\documentclass[9pt,twocolumn,twoside]{osajnl}
\usepackage{color}

\journal{pr} 

\setboolean{shortarticle}{false}

\title{Quantum non-demolition measurement based on an SU(1,1)-SU(2)-concatenated atom-light hybrid interferometer}

\author[1]{Gao-Feng Jiao}
\author[1]{Keye Zhang}
\author[1,5]{L. Q. Chen}
\author[1,6]{Chun-Hua Yuan}
\author[2,3,4]{Weiping Zhang}

\affil[1]{State Key Laboratory of Precision Spectroscopy, Quantum Institute for Light and Atoms, Department of Physics, East China Normal University, Shanghai 200062, China}
\affil[2]{School of Physics and Astronomy, and Tsung-Dao Lee Institute, Shanghai Jiao Tong University, Shanghai 200240, China}
\affil[3]{Shanghai Research Center for Quantum Sciences, Shanghai 201315, China}
\affil[4]{Collaborative Innovation Center of Extreme Optics, Shanxi University, Taiyuan, Shanxi 030006, China}
\affil[5]{Corresponding author: lqchen@phy.ecnu.edu.cn}
\affil[6]{Corresponding author: chyuan@phy.ecnu.edu.cn}

\begin{abstract}
Quantum non-demolition (QND) measurement is an important tool in the field of
quantum information processing and quantum optics. The atom-light hybrid
interferometer is of great interest due to its combination of atomic spin wave
and optical wave, which can be utilized for photon number QND measurement via the AC-Stark effect.
In this paper, we present an SU(1,1)-SU(2)-concatenated atom-light hybrid interferometer, and theoretically study the QND measurement of photon number. Compared to the traditional SU(2) interferometer, the signal-to-noise ratio (SNR) in a balanced case is improved by a gain factor of the nonlinear Raman process (NRP) in this proposed interferometer. Furthermore, the condition of high-quality of QND measurement is analyzed.
In the presence of losses, the measurement quality is reduced. We can adjust the gain parameter of the NRP in readout stage to reduce the impact due to losses. Moreover, this scheme is a multiarm interferometer, which has the potential of multiparameter estimation with many important applications in the detection of vector fields, quantum imaging and so on.

\end{abstract}

\setboolean{displaycopyright}{true}

\begin{document}

\maketitle

\section{Introduction}

Quantum measurement takes place at the interface between the quantum world
and macroscopic reality. According to quantum mechanics, as soon as we observe a system we actually perturb it, 
and such perturbation can’t be reduced to zero, instead there is a fundamental limit given by 
the Heisenberg uncertainty relation. QND is motivated to get as closer as possible to this limit \cite{1,Onofrio96,2}. In QND measurement, 
a signal observable of the measured system is coupled to
a readout observable of a probe system through an interaction, so that the information about the measured quantity can be obtained indirectly
by the direct measurement of the readout observable. The key issue is to design the measurement scheme in which 
the back action noise is transferred to the other unmeasured conjugate observable without being coupled back to the quantity of interest.
The QND measurement was studied in a variety of quantum systems,
including mechanical oscillators \cite{3,4}, trapped ions \cite{5,6},
solid-state spin qubits \cite{7,8}, circuit quantum electrodynamics \cite%
{9,10}, and photons \cite{Grangier98,11,12}.

\begin{figure*}[tb]
\centering{\includegraphics[scale=0.3,angle=0]{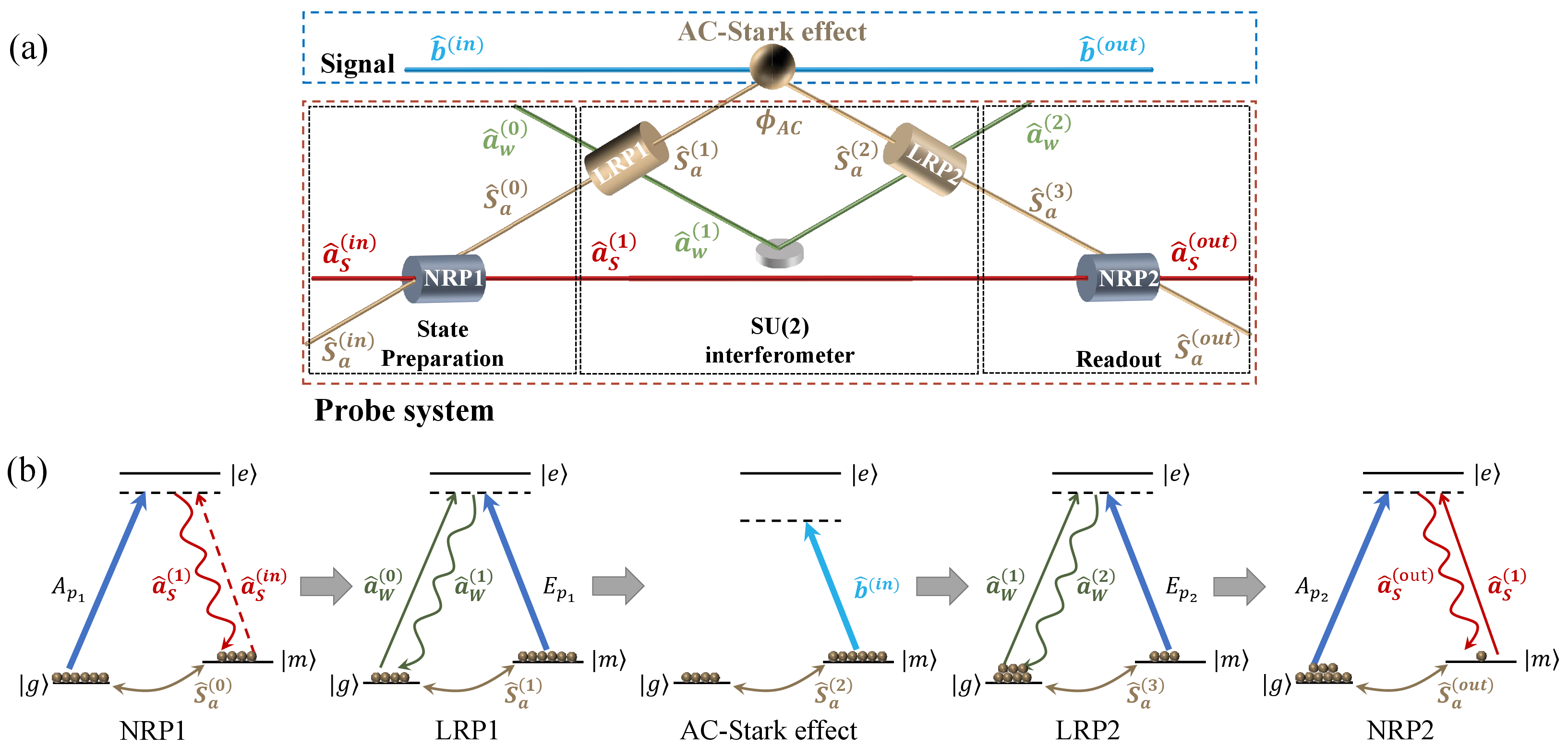}}
\caption{ (a) Schematic of QND measurement of photon number. The probe system
consists of an SU(1,1)-SU(2)-concatenated atom-light hybrid interferometer. 
In the SU(2) interferometer in the middle box, the LRP is utilized to realize the splitting and combination of
the atomic spin wave and the optical wave. $\hat{a}_{W}^{(0)}$ is a
coherent state, $\hat{S}_{a}^{(0)}$, $\hat{S}_{a}^{(1)}$ and $\hat{S}_{a}^{(3)}$ are the atomic collective excitation which
are prepared by the NRP1, LRP1 and LRP2, respectively. The atomic spin wave
$\hat{S}_{a}^{(1)}$ experiences a phase modulation $\protect\phi_{AC}$ via
the AC-Stark effect by signal light $\hat{b}^{(in)}$ and evolves to $\hat{S}_{a}^{(2)}$. The generated atomic
spin wave $\hat{S}_{a}^{(3)}$ of the SU(2) interferometer and the optical
wave $\hat{a}_{S}^{(1)}$ which is correlated with $\hat{S}_{a}^{(0)}$ are
combined to realize active correlation output readout via the NRP2. $%
\mathrm{LRP}$, linear Raman process; $\mathrm{NRP}$, nonlinear Raman
process. (b) Energy levels of the atom. The lower two energy states $|g\rangle$ and
$|m\rangle$ are the hyperfine split ground states. The higher-energy state $|e\rangle$
is the excited state. The strong pump field $A_{p_{1}}$ ($A_{p_{2}}$) and strong read field $E_{p_{1}}$ ($E_{p_{2}}$)
couple the transitions $|g\rangle$ $\rightarrow$ $|e\rangle$ and $|m\rangle$ $\rightarrow$ $|e\rangle$,
respectively. $\hat{b}^{(in)}$ is far off resonance with the transition $|m\rangle$ $\rightarrow$ $|e\rangle$
by a large detuning. }
\label{fig1}
\end{figure*}

In the quantum optics community, Imoto \textit{et al.} \cite{13} proposed an
optical interferometer for QND measurement scheme of the photon number using
the optical Kerr effect, in which the cross-Kerr interaction encodes the
photon number of the signal state onto a phase shift of the probe state.
Subsequently, some works about the photon number QND measurement are studied
\cite{14,15,16,17,44,18}. However, the main obstacles to optical QND
measurement using cross-phase modulation based on third-order nonlinear
susceptibilities $\chi^{(3)}$ are the small value of nonlinearity in the
available media and the absorption of photons. Munro \textit{et al.} \cite%
{19} presented an implementation of the QND measurement scheme with the
required nonlinearity provided by the giant Kerr effect achievable with
AC-Stark electromagnetically transparency. In addition, the schemes using
cavity or circuit quantum electrodynamical systems have enabled photon
number QND measurements that do not rely on the material nonlinearity
through strong light-matter interactions \cite{20,21,22,23,24}.

Interferometric measurements allow for highly sensitive detection of any small changes 
that induces an optical phase shift. Different kinds of quantum interferometers have been proposed 
for phase measurement \cite{45,46,47,48,49,50,51,52,53,54}.
The phase sensitivity of the SU(2) interferometer with coherent state input, such as the Michelson or 
Mach-Zehnder interferometer, is limited by the shot noise limit.
When the unused port of the interferometer is fed with squeezed state, the SU(2) interferometer can beat this limit \cite{45}.
In contrast to the SU(2) interferometer, the SU(1,1) nonlinear interferometer that replaces the beam splitters in the SU(2) interferometer with
nonlinear gain media is a fundamentally different type of interferometric detector \cite{46}, 
in which the quantum correlations are generated within the interferometer,
the phase sensitivity can, in principle, approach the Heisenberg limit. 

Recently, using the coherent mixing of an optical wave and an atomic spin
wave, the atom-light hybrid interferometer has been proposed and studied \cite{25,26,27,28,29,30,56}. 
Similarly, there are two types of atom-light hybrid
interferometers, which have been also demonstrated experimentally. One is
SU(2)-type atom-light hybrid interferometer \cite{27}, where the linear Raman processes
(LRPs) replace the beam splitters in the SU(2) linear interferometer to
realize linear superposition of atomic wave and optical wave. The
other is SU(1,1)-type atom-light hybrid interferometer \cite{28}, where the nonlinear
Raman processes (NRPs) realize the atomic wave and optical wave splitting and
recombination. The atom-light hybrid interferometer has drawn
considerable interest because it is sensitive to both optical and atomic phase
shift. Resorting to an AC Stark effect for the atomic phase change, the
atom-light hybrid interferometer can be applied to QND measurement of photon
number \cite{31,32}. Further developing methods to improve the measurement
procision is always desired for the innovation of interferometers.

In this paper, we present an SU(1,1)-SU(2)-concatenated atom-light hybrid
interferometer, which is combination of SU(2) and SU(1,1) type atom-light
hybrid interferometers. Then the AC-Stark effect encodes the photon number of
the signal light onto a phase shift of the atomic spin wave, so that the
problem of detecting photon number is translated into the problem of detecting
the atomic phase shift. Thus we theoretically analyze the SNR of the
interferometer for precision measurement of the phase shift to examine the
performance of this scheme as a QND measurement. Besides, an ideal QND
measurement can be testified by a perfect correlation between the signal light
and the probe system \cite{33}. Here we estimate the quality of QND
measurement using the criteria in Ref.~\cite{33}, and the condition for a
perfect correlation is given. In the presence of losses, the measurement
quality is reduced. However, we can adjust the gain parameter of the NRP in
readout stage to reduce the impact due to losses. 

Our scheme can be thought of as inserting an SU(2)-type atom-light hybrid interferometer into one of the arms of the SU(1,1)-type atom-light hybrid interferometer. Compared to a conventional SU(1,1) interferometer, the number of phase-sensing particles is further increased due to input field $\hat{a}_{W}^{(0)}$. In the previous scheme \cite{31}, a strong stimulated Raman process can indeed be used to excite most of the atoms to the
$|m\rangle$ state to improve the number of phase-sensing particles. However,
this will produce additional nonlinear effects and noise light fields
\cite{41,42,43}. Therefore, in the current scheme, firstly a spontaneous Raman
process is used, and then a LRP is used to generate the Rabi-like
superposition oscillation between light and the atom. By controlling the
interaction time, more atoms can be prepared to the $|m\rangle$ state to
improve the number of phase-sensing particles to enhance the QND measurement.
Compared to the previous works \cite{31,32}, the SU(1,1)-SU(2)-concatenated
atom-light hybrid interferometer can be realized QND measurement of photon
number with higher precision. Since the SU(2)-type and SU(1,1)-type atom-light hybrid interferometers \cite{27,28} have been demonstrated experimentally, the SU(1,1)-SU(2)-concatenated atom-light hybrid
interferometer could be realized with current experimental conditions and the corresponding experimental requirements are given in the discussion part. 

\section{An SU(1,1)-SU(2)-concatenated Atom-light hybrid interferometer}

An SU(1,1)-SU(2)-concatenated atom-light hybrid interferometer is shown in Fig.~\ref{fig1} (a). 
The corresponding energy levels of atom and optical
frequencies for the formation of the SU(1,1)-SU(2)-concatenated
atom-light hybrid interferometer are given in Fig.~\ref{fig1} (b), where the lower two energy states $|g\rangle$ and
$|m\rangle$ are the hyperfine split ground states. The higher-energy state $|e\rangle$
is the excited state. The strong pump field $A_{p_{1}}$ ($A_{p_{2}}$) and strong read field $E_{p_{1}}$ ($E_{p_{2}}$)
couple the transitions $|g\rangle$ $\rightarrow$ $|e\rangle$ and $|m\rangle$ $\rightarrow$ $|e\rangle$,
respectively. The process can be used to describe\ the linear and
nonlinear splitting and recombination of the light field and spin wave, i.e.,
NRP and LRP.

In the case of undepleted read field approximation,
the two-mode coupled Hamiltonian is written as \cite{27}%
\begin{equation}
\hat{H}_{LRP}=i\hbar\Omega\hat{a}_{W}\hat{S}_{a}^{\dagger}+H.c.,
\end{equation}
where $\hat{S}_{a}\equiv(1/\sqrt{N})\sum_{k}|g\rangle_{kk}\langle
m|$ is the spin wave (atomic collective
excitation) with $N$\ the number of atoms in the ensemble, and
$\Omega$ is the Rabi-like frequency. The Rabi-like oscillation between the write field and the atomic spin
wave occurs, and the input-output relation is given by%
\begin{align}
\hat{S}_{a}^{(out)}=\hat{S}_{a}^{(in)}cos(|\Omega|\tau_{r})-\hat{a}_{W}^{(in)}sin(|\Omega|\tau_{r}), \nonumber\\
\hat{a}_{W}^{(out)}=\hat{a}_{W}^{(in)}cos(|\Omega|\tau_{r})+\hat{S}_{a}^{(in)}sin(|\Omega|\tau_{r}), \label{eq11}
\end{align}
where $\tau_{r}$ is the interaction time of read field $E_{p}$. This
transform of Eq.~(\ref{eq11}) can be used for LRP. In the case of undepleted pump field approximation, the 
Hamiltonian is written as \cite{28}
\begin{equation}
\hat{H}_{NRP}=i\hbar{\eta}A_{p}\hat{a}_{S}^{\dag}\hat{S}_{a}^{\dag}+H.c.,
\end{equation}
where $\eta$ is the coupling coefficient, and $A_{p}$ is the
amplitude of the pump beam. The input-output
relation is given by%
\begin{equation}
\hat{S}_{a}^{(out)}=G\hat{S}_{a}^{(in)}+ge^{i\theta}\hat{a}_{S}^{\dagger
(in)},\text{ }\hat{a}_{S}^{(out)}=G\hat{a}_{S}^{(in)}+ge^{i\theta}\hat{S}_{a}%
^{\dagger(in)}, \label{eq10}
\end{equation}
where $e^{i\theta}=\eta A_{p}/|\eta A_{p}|$, $G%
=\cosh(|\eta A_{p}|\tau_{p})$ and $g=\sinh(|\eta A_{p}|\tau_{p})$ are the
gains of the process with $G^{2}-g^{2}=1$, $\tau_{p}$ the pulse duration
of pump beam $A_{p}$. This transform of Eq.~(\ref{eq10})
can be used for NRP. 

Next, according to above linear and nonlinear transforms, the output operators
of the SU(1,1)-SU(2)-concatenated atom-light hybrid interferometer is worked out as a function of the input operators with
three steps: (1) the state preparation; (2) the SU(2) interferometer, (3) the
readout. In the first step, we prepare the initial atomic spin wave $\hat
{S}_{a}^{(0)}$ and a correlated optical wave $\hat{a}_{S}^{(1)}$ via the NRP1.
It can be described as%
\begin{equation}
\hat{S}_{a}^{(0)}=G_{1}\hat{S}_{a}^{(in)}+g_{1}e^{i\theta_{1}}\hat{a}%
_{S}^{\dagger(in)},\text{ }\hat{a}_{S}^{(1)}=G_{1}\hat{a}_{S}^{(in)}%
+g_{1}e^{i\theta_{1}}\hat{S}_{a}^{\dagger(in)}.
\end{equation}

Next, the beam splitter in a SU(2) interferometer is provided by the LRP,
which can split and mix the atomic spin wave and the optical wave coherently
for interference. $\hat{a}_{W}^{(0)}$ is a coherent
state, $\hat{S}_{a}^{(0)}$ is an atomic collective excitation which is
prepared by the NRP1. The relationship between input and output in the SU(2)
interferometer is given by%
\begin{equation}
\hat{S}_{a}^{(3)}=t\hat{S}_{a}^{(0)}+r\hat{a}_{W}^{(0)},\text{ }\hat{a}%
_{W}^{(2)}=t\hat{a}_{W}^{(0)}+r\hat{S}_{a}^{(0)}, \label{eq12}
\end{equation}
where $t=e^{i\phi/2}\cos(\phi/2)$, $r=ie^{i\phi/2}\sin(\phi/2)$. $\phi$ denotes the phase shift.

In the final step, the generated atomic spin wave $\hat{S}_{a}^{(3)}$ of the
SU(2) interferometer and the optical wave $\hat{a}_{S}^{(1)}$ are combined to
realize active correlation output readout via the NRP2. It can be expressed as%
\begin{equation}
\hat{S}_{a}^{(out)}=G_{2}\hat{S}_{a}^{(3)}+g_{2}e^{i\theta_{2}}\hat{a}%
_{S}^{\dagger(1)},\text{ }\hat{a}_{S}^{(out)}=G_{2}\hat{a}_{S}^{(1)}%
+g_{2}e^{i\theta_{2}}\hat{S}_{a}^{\dagger(3)}.
\end{equation}

And thus, the full input-output relation of the actively correlated atom-light
hybrid interferometer is%
\begin{align}
\hat{a}_{S}^{(out)}  &  =A\hat{a}_{S}^{(in)}+B\hat{S}_{a}^{\dagger(in)}%
+C\hat{a}_{W}^{\dagger(0)},\nonumber\\
\hat{S}_{a}^{(out)}  &  =D\hat{a}_{S}^{\dagger(in)}+E\hat{S}_{a}^{(in)}%
+F\hat{a}_{W}^{(0)},
\end{align}
where%
\begin{align}
A  &  =G_{2}G_{1}+g_{2}g_{1}e^{i(\theta_{2}-\theta_{1})}t^{\ast},\text{ }B=G_{2}g_{1}e^{i\theta_{1}}+G_{1}g_{2}e^{i\theta_{2}}t^{\ast},\nonumber\\
D  &  =G_{1}g_{2}e^{i\theta_{2}}+G_{2}g_{1}e^{i\theta_{1}}t,\text{ }E=g_{2}g_{1}e^{i(\theta_{2}-\theta_{1})}+G_{2}G_{1}t,\nonumber\\
C  &  =g_{2}e^{i\theta_{2}}r^{\ast},\text{ }F=G_{2}r.
\end{align}

\section{QND measurement of photon number}

This new type interferometer in Section 2 can be used as a probe system for QND measurement. The schematic of QND measurement of photon number is shown in Fig.~\ref{fig1} (a),
when the atoms system of this new type interferometer are illuminated by the off-resonant signal light $\hat{b}^{(in)}$,  the interaction between atom and signal light $\hat{b}^{(in)}$ will
induce an atomic phase shift  \cite{34,40}
\begin{align} 
\phi_{AC}=\kappa\hat{n}_{b}, 
\end{align}
where $\hat{n}_{b}=\hat{b}^{(in)\dag}\hat{b}^{(in)}$ is photon number operator of the signal light and
$\kappa$ is the AC-Stark coefficient.

In this QND measurement scheme, the photon number of the signal light is the QND observable, 
while the phase of the signal light is the conjugated observable. The quantum noise induced by the act of measurement 
is fed into the unmeasured conjugate observable $\phi^{(s)}$ of the signal light, where superscript $(s)$ denotes the signal light.
The uncertainty of these two observables is limited by the Heisenberg relation $\Delta n_{b}^{(s)} \Delta\phi^{(s)} \ge 1$ \cite{Heitler}.
That is, the QND measurement of the observable $n_{b}^{(s)}$ is accomplished at the expense of uncertainty increasement 
of its conjugate observable $\phi^{(s)}$.
By monitoring the atomic phase shift $\phi_{AC}$ using this atom-light hybrid interferometer, 
we can determine the photon number of the signal light without destroying the photons. The phase shift of the atomic spin wave is the readout observable, 
which can be measured by the homodyne detection. Next, we analyze the performance of this scheme as a QND measurement.

\subsection{SNR analysis}

The QND measurement scheme that uses the AC-Stark effect is based on
the precision measurement of the photon-induced atomic phase shift, thus the SNR analysis is helpful to examine the performance of
the QND measurement process. Given the homodyne detection, the SNR is defined
as%
\begin{equation}
R=\frac{\langle\hat{X}_{S}^{(out)}\rangle^{2}}{\langle\Delta^{2}\hat{X}_{S}^{(out)}\rangle%
},\label{eq1}%
\end{equation}
where $\langle\hat{X}_{S}^{(out)}\rangle$ and $\langle\Delta^{2}\hat{X}_{S}^{(out)}\rangle$
denote the quantum expectation and variance of the amplitude quadrature,
respectively. In our scheme, they are given by%
\begin{align}
\langle\hat{X}_{S}^{(out)}\rangle &  =\langle\hat{a}_{S}^{(out)}+\hat{a}%
_{S}^{\dagger(out)}\rangle\nonumber\\
&  =g_{2}N_{\alpha}^{1/2}[\cos(\theta_{2}-\theta_{\alpha}-\phi)-\cos(\theta_{2}%
-\theta_{\alpha})].\label{eq2}%
\end{align}%
\begin{align}
\langle\Delta^{2}\hat{X}_{S}^{(out)}\rangle &  =G_{2}^{2}G_{1}^{2}+G_{2}^{2}g_{1}^{2}%
+g_{2}^{2}(1-\cos\phi)/2\nonumber\\
&  +g_{2}^{2}g_{1}^{2}(1+\cos\phi)/2+G_{1}^{2}g_{2}^{2}(1+\cos\phi
)/2\nonumber\\
&  +2G_{2}G_{1}g_{2}g_{1}\cos(\theta_{2}-\theta_{1}-\phi)\nonumber\\
&  +2G_{2}G_{1}g_{2}g_{1}\cos(\theta_{2}-\theta_{1}).\label{eq3}%
\end{align}
Here $\hat{a}_{S}^{(in)}$ and $\hat{S}_{a}^{(in)}$ are in vacuum states,
$\hat{a}_{W}^{(0)}$ is in a coherent state $|\alpha\rangle$ with
$\alpha=N_{\alpha}^{1/2}e^{i\theta_{\alpha}}$ where $N_{\alpha}$ and $\theta_{\alpha}$ are the photon number and initial
phase of the coherent state, respectively. The phase shift $\phi$ includes the phase difference $\phi_{0}$ of the interferometer and the phase
difference $\phi_{AC}$ caused by the AC-Stark effect with $\phi_{AC}=\kappa\hat{n}_{b}$.
Our scheme can be thought of as inserting a SU(2) interferometer into one of the arms of the SU(1,1) interferometer.
For a balanced SU(1,1) interferometer configuration, two NRPs of equal gain ($g_{2}=g_{1}=g$) and opposite pump phases ($\theta_{1}=0$, $\theta_{2}=\pi$) 
are arranged in series. NRP1 produces an optical field together with a correlated atomic spin excitation, while NRP2 is shifted in pump phase to exactly 
reverse the operation of NRP1 and return the optical field and atomic spin excitation back to their original input states. When a phase shift is 
introduced, the transfer is no longer complete, and thus lead to a change in the output corresponding to the induced phase shift. Here we first consider 
a balanced case in which the SU(2) interferometer is introduced into a balanced SU(1,1) interferometer. With a small $\phi_{AC}$ around $\phi_{0}=0$ 
and $\alpha=iN_{\alpha}^{1/2}$, assuming the signal light is in a number state $|n\rangle$ with $\hat{n}_{b}|n\rangle=n_{b}|n\rangle$ 
and substituting $\phi_{AC}$ with $\kappa\hat{n}_{b}$, the SNR is
\begin{equation}
R\approx g^{2}\kappa^{2}N_{\alpha}n_{b}^{2}.
\end{equation}
To have single-photon resolution, we need AC-Stark coefficient $\kappa\sim1/g N_{\alpha}^{1/2}$. Compared to
the traditional SU(2) interferometer \cite{31}, the required coefficient is smaller by a
factor of $1/g$ . This is due to the method of active correlation
output readout for the SU(2) interferometer \cite{35}, in which the SNR is 
greater than the SU(2) interferometer by a factor of $g$ of NRP.

\subsection{Quality estimation}

In the QND measurement, the signal light
is coupled to the probe system, and a subsequent readout measurement 
of the probe output is made to extract the information about the signal light without perturbing it.
An ideal QND measurement requires a perfect correlation between the signal and the probe system.
In practice, the probe output itself has fluctuation, which leads to a nonideal QND measurement.
Thus the quality of the QND measurement scheme is worth studying.
In this section, we estimate the quality using the criteria
introduced by Holland \textit{et al.} \cite{33}, which are%
\begin{align}
C_{S^{in}S^{out}}^{2}& =\frac{|\langle S^{in}S^{out}\rangle -\langle
S^{in}\rangle \langle S^{out}\rangle |^{2}}{\langle \Delta ^{2}S^{in} \rangle\langle \Delta
^{2}S^{out}\rangle},  \label{eq4} \\
C_{S^{in}P^{out}}^{2}& =\frac{|\langle S^{in}P^{out}\rangle -\langle
S^{in}\rangle \langle P^{out}\rangle |^{2}}{\langle \Delta ^{2}S^{in} \rangle\langle  \Delta
^{2}P^{out}\rangle},  \label{eq5} \\
C_{S^{out}P^{out}}^{2}& =\frac{|\langle S^{out}P^{out}\rangle -\langle
S^{out}\rangle \langle P^{out}\rangle |^{2}}{\langle \Delta ^{2}S^{out} \rangle\langle \Delta
^{2}P^{out}\rangle},  \label{eq6}
\end{align}%
with%
\begin{align}
\langle \Delta ^{2}S^{in} \rangle& =\langle (S^{in})^{2}\rangle -\langle S^{in}\rangle ^{2},
\notag \\
\langle \Delta ^{2}S^{out} \rangle& =\langle (S^{out})^{2}\rangle -\langle S^{out}\rangle
^{2},  \notag \\
\langle \Delta ^{2}P^{in} \rangle& =\langle (P^{in})^{2}\rangle -\langle P^{in}\rangle ^{2},
\notag \\
\langle \Delta ^{2}P^{out} \rangle& =\langle (P^{out})^{2}\rangle -\langle P^{out}\rangle
^{2},
\end{align}%
where $S^{in}$ is the input signal incident on the scheme and $P^{out}$ is
the output probe measured by a detector. Here $S^{in}$ is the photon number
of the input signal $\hat{N}^{(in)}=\hat{b}^{\dagger (in)}\hat{b}^{(in)}$,
and the probe is the amplitude quadrature $\hat{X}_{S}^{(out)}$. Eq.~(\ref{eq4}) is about how much the probe system degrades the signal of the measured
system. Eq.~(\ref{eq5}) is about how good the probe system is a measurement
device. Eq.~(\ref{eq6}) is that how good the probe system is a state preparation
device. For a ideal QND measurement device the correlation coefficients $%
C_{S^{in}S^{out}}^{2}$, $C_{S^{in}P^{out}}^{2}$ and $C_{S^{out}P^{out}}^{2}$
are unity. In our paper, the signal light leading to the AC-Stark shift is
far off resonance with a large detuning, i.e., the photon number of the
measured signal light is not changed before and after measurement. So the
first criterion is satisfied, the second and the third criteria become same:
$C_{\hat{N}^{(in)}\hat{X}_{S}^{(out)}}^{2}=C_{\hat{N}^{(out)}\hat{X}%
_{S}^{(out)}}^{2}$, that is,%
\begin{equation}
C^{2}=\frac{|\langle \hat{N}^{(in)}\hat{X}_{S}^{(out)}\rangle -\langle \hat{N%
}^{(in)}\rangle \langle \hat{X}_{S}^{(out)}\rangle |^{2}}{\langle \Delta ^{2}\hat{N}%
^{(in)}\rangle \langle \Delta ^{2}\hat{X}_{S}^{(out)}\rangle}.  \label{eq7}
\end{equation}%
For brevity, we omit the subscript of $C$. In the previous section, even if we assume the signal light is in a number state,
which is the eigen-state of the photon number measurement process ($\hat{n}_{b}|n\rangle=n_{b}|n\rangle$).
The probe output has fluctuation and does not yield a definite value. Here, for convenience, the signal light is set in a 
coherent state $|\beta \rangle $ with photon number $N_{\beta }$.
We obtain%
\begin{align}
\langle \hat{N}^{(in)}\rangle & =N_{\beta },\text{ }\Delta ^{2}\hat{N}%
^{(in)}=N_{\beta },  \notag \\
\langle \hat{X}_{S}^{(out)}\rangle & =g_{2}\kappa N_{\alpha }^{1/2}N_{\beta
},\text{ }\langle \hat{N}^{(in)}\hat{X}_{S}^{(out)}\rangle =g_{2}\kappa N_{\alpha
}^{1/2}N_{\beta }(N_{\beta }+1),  \notag \\
\langle \Delta ^{2}\hat{X}_{S}^{(out)}\rangle&
=(G_{2}G_{1}-g_{2}g_{1})^{2}+(G_{2}g_{1}-G_{1}g_{2})^{2}  \notag \\
& +g_{2}^{2}\kappa ^{2}N_{\alpha }N_{\beta }+G_{1}^{2}g_{2}^{2}\kappa
^{2}N_{\beta }(N_{\beta }+1)/2.  \label{eq8}
\end{align}%
Substituting Eq.~(\ref{eq8}) into Eq.~(\ref{eq7}), the
criteria can be written as%
\begin{equation}
C^{2}=\frac{1}{1+\frac{%
(G_{2}G_{1}-g_{2}g_{1})^{2}+(G_{2}g_{1}-G_{1}g_{2})^{2}}{g_{2}^{2}\kappa
^{2}N_{\alpha }N_{\beta }}+\frac{G_{1}^{2}(N_{\beta }+1)}{2N_{\alpha }}}.
\label{eq9}
\end{equation}%
As seen in Eq.~(\ref{eq9}), a perfect correlation $C^{2}\approx 1$ can be
satisfied under the condition of $g_{2}^{2}\kappa ^{2}N_{\alpha }N_{\beta
}\gg (G_{2}G_{1}-g_{2}g_{1})^{2}+(G_{2}g_{1}-G_{1}g_{2})^{2}$ and $%
2N_{\alpha }\gg G_{1}^{2}(N_{\beta }+1)$. In a balanced case, the condition
for a perfect correlation is $g^{2}\kappa ^{2}N_{\alpha }N_{\beta }\gg 1$
and $2N_{\alpha }\gg G^{2}(N_{\beta }+1)$.

\begin{figure}[t]
\centering{\includegraphics[scale=0.45,angle=0]{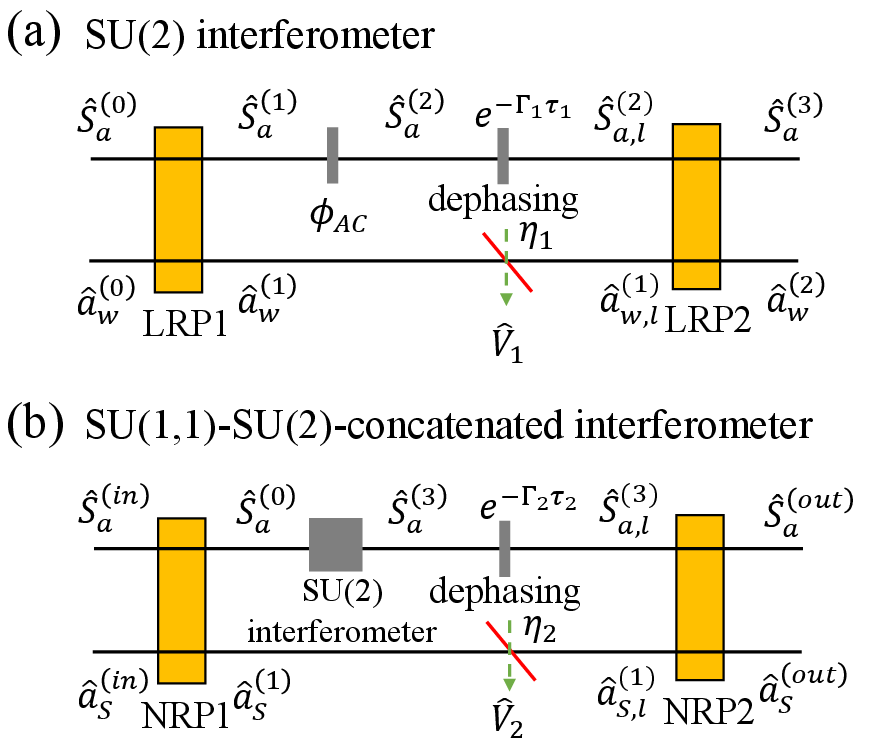}}\caption{ A lossy interferometer model with (a) internal loss and (b) external loss.}%
\label{fig3}%
\end{figure}

\subsection{Optimized C in the presence of losses}

Next, we investigate the effects of losses on the correlation coefficient $C$
in the presence of photon losses and atomic decoherence losses \cite{30}.

The loss of the two arms inside the SU(2) interferometer is called the internal loss, as shown in Fig.~\ref{fig3}(a). The loss at the output of SU(2) and the associated optical field is called external loss, as shown in Fig.~\ref{fig3}(b). Two fictitious beam splitters are introduced to mimic the loss of photons into the
environment, then the optical waves $\hat{a}%
_{W}^{(1)}$ and $\hat{a}_{S}^{(1)}$ experience losses as%
\begin{align}
\hat{a}_{W,l}^{(1)} &  =\sqrt{\eta_{1}}\hat{a}_{W}^{(1)}+\sqrt{1-\eta_{1}%
}\hat{V}_{1},\\
\hat{a}_{S,l}^{(1)} &  =\sqrt{\eta_{2}}\hat{a}_{S}^{(1)}+\sqrt{1-\eta_{2}%
}\hat{V}_{2},
\end{align}
where subscript $l$ indicates the loss, $\eta_{1}$ ($\eta_{2}$) and $\hat{V}_{1}$ ($\hat{V}_{2}$) represent the
transmission rate and vacuum, respectively. The spin waves $\hat{S}_{a}^{(2)}$ ($\hat{S}_{a}^{(3)}$) also undergoes the collisional dephasing $e^{-\Gamma
_{1}\tau_{1}}$ ($e^{-\Gamma_{2}\tau_{2}}$), then the spin waves are described
by%
\begin{align}
\hat{S}_{a,l}^{(2)} &  =\hat{S}_{a}^{(2)}e^{-\Gamma_{1}\tau_{1}}+\hat{F}%
_{1},\\
\hat{S}_{a,l}^{(3)} &  =\hat{S}_{a}^{(3)}e^{-\Gamma_{2}\tau_{2}}+\hat{F}%
_{2},
\end{align}
where $\langle\hat{F}_{1}\hat{F}_{1}^{\dagger}\rangle=1-e^{-2\Gamma_{1}%
\tau_{1}}$ and $\langle\hat{F}_{2}\hat{F}_{2}^{\dagger}\rangle=1-e^{-2\Gamma
_{2}\tau_{2}}$ guarantees the consistency of the operator properties of
$\hat{S}_{a,l}^{(2)}$ and $\hat{S}_{a,l}^{(3)}$, respectively. The
input-output relation for loss case of $\hat{a}_{S}^{(out)}$ becomes%
\begin{align}
\hat{a}_{S,l}^{(out)}  & =\hat{a}_{S}^{(in)}\mathcal{A}+\hat{S}_{a}%
^{\dagger(in)}\mathcal{B}+\hat{a}_{W}^{\dagger(0)}\mathcal{C}+\hat{V}%
_{1}^{\dagger}\mathcal{D}+\hat{V}_{2}\mathcal{E}\nonumber\\
& +\hat{F}_{1}^{\dagger}\mathcal{F}+\hat{F}_{2}^{\dagger}\mathcal{G},
\end{align}
where%
\begin{align}
\mathcal{A} &  =[\sqrt{\eta_{2}}G_{2}G_{1}+g_{2}g_{1}e^{i(\theta_{2}%
-\theta_{1})}(e^{-\Gamma_{1}\tau_{1}}e^{-i\phi}+\sqrt{\eta_{1}})e^{-\Gamma
_{2}\tau_{2}}/2],\nonumber\\
\mathcal{B} &  =[\sqrt{\eta_{2}}G_{2}g_{1}e^{i\theta_{1}}+G_{1}g_{2}%
e^{i\theta_{2}}(e^{-\Gamma_{1}\tau_{1}}e^{-i\phi}+\sqrt{\eta_{1}}%
)e^{-\Gamma_{2}\tau_{2}}/2],\nonumber\\
\mathcal{C} &  =g_{2}e^{i\theta_{2}}(e^{-\Gamma_{1}\tau_{1}}e^{-i\phi}%
-\sqrt{\eta_{1}})e^{-\Gamma_{2}\tau_{2}}/2,\nonumber\\
\mathcal{D} &  =-g_{2}e^{i\theta_{2}}\sqrt{1-\eta_{1}}e^{-\Gamma_{2}\tau_{2}%
}/\sqrt{2},\nonumber\\
\mathcal{E} &  =G_{2}\sqrt{1-\eta_{2}},\mathcal{F}=g_{2}e^{i\theta_{2}%
}e^{-\Gamma_{2}\tau_{2}}/\sqrt{2},\mathcal{G}=g_{2}e^{i\theta_{2}}.
\end{align}

We study the effect of losses under the condition of $\theta_{1}=0$,
$\theta_{2}=\pi$, $\theta_{\alpha}=\pi/2$, and $\phi_{0}=0$ with a small
$\phi_{AC}$ around $\phi_{0}$. Considering losses the terms of $\langle\hat
{X}_{S}^{(out)}\rangle_{l}$, $\langle\hat{N}^{(in)}\hat{X}_{S}^{(out)}%
\rangle_{l}$, and $\langle\Delta^{2}\hat{X}_{S}^{(out)}\rangle_{l}$ in Eq.~(\ref{eq8}) are
given by
\begin{equation}
\langle\hat{X}_{S}^{(out)}\rangle_{l}=g_{2}e^{-\Gamma_{1}\tau_{1}}%
e^{-\Gamma_{2}\tau_{2}}\kappa N_{\alpha}^{1/2}N_{\beta},
\end{equation}%
\begin{equation}
\langle\hat{N}^{(in)}\hat{X}_{S}^{(out)}\rangle_{l}=g_{2}e^{-\Gamma_{1}%
\tau_{1}}e^{-\Gamma_{2}\tau_{2}}\kappa N_{\alpha}^{1/2}N_{\beta}(N_{\beta}+1),
\end{equation}
and%
\begin{align}
&  \langle\Delta^{2}\hat{X}_{S}^{(out)}\rangle_{l}\nonumber\\
&  =(\sqrt{\eta_{2}}G_{2}G_{1}-g_{2}g_{1}e^{-\Gamma_{1}\tau_{1}}e^{-\Gamma
_{2}\tau_{2}}/2-g_{2}g_{1}e^{-\Gamma_{2}\tau_{2}}\sqrt{\eta_{1}}%
/2)^{2}\nonumber\\
&  +(\sqrt{\eta_{2}}G_{2}g_{1}-G_{1}g_{2}e^{-\Gamma_{1}\tau_{1}}e^{-\Gamma
_{2}\tau_{2}}/2-G_{1}g_{2}e^{-\Gamma_{2}\tau_{2}}\sqrt{\eta_{1}}%
/2)^{2}\nonumber\\
&  +g_{2}^{2}[(1-e^{-2\Gamma_{1}\tau_{1}})e^{-2\Gamma_{2}\tau_{2}%
}/2+(1-e^{-2\Gamma_{2}\tau_{2}})]\nonumber\\
&  +g_{2}^{2}(2g_{1}^{2}+1)\kappa^{2}N_{\beta}(N_{\beta}+1)e^{-2\Gamma_{1}%
\tau_{1}}e^{-2\Gamma_{2}\tau_{2}}/4\nonumber\\
&  +g_{2}^{2}\kappa^{2}N_{\beta}[N_{\alpha}+(N_{\beta}+1)/4]e^{-2\Gamma
_{1}\tau_{1}}e^{-2\Gamma_{2}\tau_{2}}\nonumber\\
&  +(g_{2}e^{-\Gamma_{2}\tau_{2}}\sqrt{\eta_{1}}/2-g_{2}e^{-\Gamma_{1}\tau
_{1}}e^{-\Gamma_{2}\tau_{2}}/2)^{2}\nonumber\\
&  +g_{2}^{2}(1-\eta_{1})e^{-2\Gamma_{2}\tau_{2}}/2+G_{2}^{2}(1-\eta_{2}),
\end{align}
then the QND measurement criterion for loss case can be obtained according to
Eq.~(\ref{eq7}).

\begin{figure}[t]
\centering{\includegraphics[scale=0.55,angle=0]{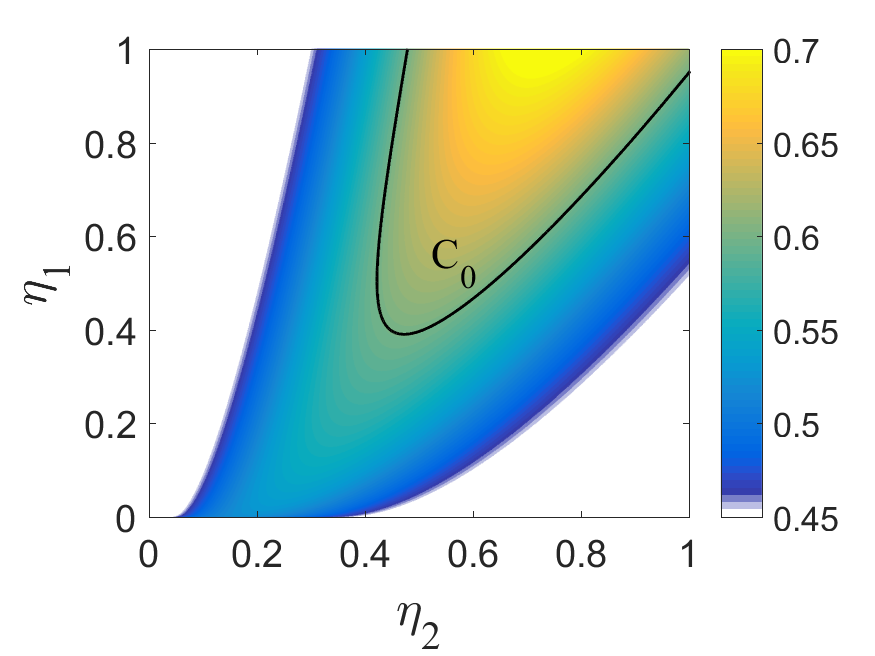}}\caption{ Correlation coefficient $C$ as a function of $\eta_1$ and $\eta_2$, where $e^{-\Gamma_{1}\tau_{1}}=e^{-\Gamma_{2}\tau_{2}}=0.9$, $\kappa=10^{-10}%
$, \text{} $g_{1}=g_{2}=3$, $N_{\alpha}=10^{12}$, \text{} and $N_{\beta}=10^{8}$. The correlation coefficient in the area of upper right corner and within the $C_{0}$ lines can be kept above 0.6.}%
\label{fig4}%
\end{figure}

In our scheme, the atomic spin wave stays in the atomic ensemble while the
optical field travels out of the atomic ensemble. Here, within the coherence
time the atomic collisional dephasing loss $\Gamma_{1}\tau_{1}$ ($\Gamma
_{2}\tau_{2}$) is small, then we set $e^{-\Gamma_{1}\tau_{1}}%
=e^{-\Gamma_{2}\tau_{2}}=0.9$. The correlation coefficient $C$ as a function
of $\eta_{1}$ and $\eta_{2}$ in balanced case is shown in Fig.~\ref{fig4}. It is shown that the reduction in the correlation coefficient $C$ increases as
the loss increases. It is shown that
our scheme is more tolerant with the internal photon loss $\eta_1$ compared to the
photon loss outside the SU(2) interferometer $\eta_2$. The reason behind the
phenomenon is that large external photon loss affects quantum correlation
between the light wave and atomic spin wave, which destroys the active
correlation output readout.  

\begin{figure}[t]
\centering{\includegraphics[scale=0.55,angle=0]{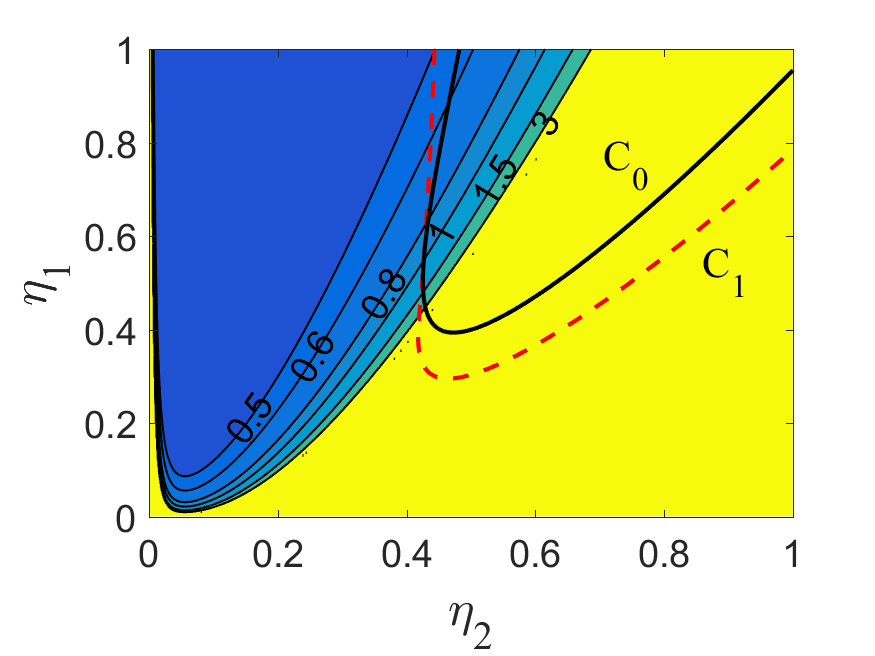}}
\caption{ Contour line of optimized
$g_{2}/g_{1}$ as a function of $\eta_{1}$ and $\eta_{2}$ with $e^{-\Gamma
_{1}\tau_{1}}=e^{-\Gamma_{2}\tau_{2}}=0.9$, \text{} where $\kappa=10^{-10}$, \text{} $g_{1}%
=3$, \text{} $N_{\alpha}=10^{12}$, \text{} and $N_{\beta}=10^{8}$. The correlation coefficient
in the area of upper right corner and within the line $C_{0}$ (before optimization) or $C_{1}$ (after optimization) can
be kept above $0.6$.}%
\label{fig5}%
\end{figure}

In the unbalanced case ($g_{1}\neq g_{2}$), we can adjust the gain ratio
$g_{2}/g_{1}$ of the beam recombination process to reduce the reduction in
correlation coefficient. The black line in Fig.~\ref{fig4} is labeled as $C_{0}$, as a benchmark for comparison 
before and after optimization, here we set $C_{0}=0.6$. The correlation coefficient in the area of upper right
corner and within the $C_{0}$ lines can be kept above $0.6$. After optimizing the $g_{2}$, the correlation coefficient as a
function of $\eta_{1}$ and $\eta_{2}$ can also be obtained, where the line
with $C$ equal to $0.6$ is denoted as $C_{1}$. The contour line of optimized
$g_{2}/g_{1}$ as a function of $\eta_{1}$ and $\eta_{2}$ are shown in Fig.
\ref{fig5}, where the position of $C_{0}$ (before optimization) and $C_{1}$ (after optimization) in the contour figure of the correlation coefficient versus
transmission rates has changed. It is demonstrated that for a given $g_{1}$
by optimizing $g_{2}/g_{1}$, the small area between the $C_{0}$ and $C_{1}$
can still kept above $0.6$. That is within a certain losses range, $C$ can
continue to beat the criteria (such as $C$ equal to $0.6)$ after optimizing
$g_{2}/g_{1}$. The newly added area after optimization is divided into two parts, one of which is the ratio $g_{2}/g_{1}$ within the very small area above is less than 1, and the other part the ratio $g_{2}/g_{1}$ needs to be greater than 1.

\section{Discussion and Conclusion}

In our scheme, the spin wave prepared by the first NRP participates in the
subsequent LRP and NRP and additional operations are required within the
coherence time of the spin wave compared to previous experimental works on atom-light
interferometers \cite{27,28}, therefore the pulse width and intensity of the
pump field $A_{p_{1}}$ ($A_{p_{2}}$) and the read field $E_{p_{1}}$ ($E_{p_{2}}$) should be properly arranged. 
Experimental consideration of the implementation of the scheme may be
performed with a rubidium atomic vapor in a cell. The
energy levels of the Rb atom are shown in Fig.~\ref{fig1} (b), where states $|g\rangle $ and $|m\rangle $ are 
the two ground states $|5^{2}S_{1/2},F=1,2\rangle $ from hyperfine splitting
and $|e\rangle$ is the excited state $|5^{2}P_{1/2},F=2\rangle $. The signal field is far resonance with the
transition $|5^{2}S_{1/2},F=2\rangle $ $\rightarrow$ $|5^{2}P_{1/2},F=2\rangle $
by $2-4$ GHz detuning \cite{27}. The interaction between atom and
signal light will induce an atomic phase shift which is proportional to the
photon number of the signal field. With a detuning 2 GHz and photon number of signal light $N_{\beta}=10^{8}$, the AC Stark coupling coefficient $\kappa$ is  $\sim10^{-10}$ rad per photon \cite{31}. With a gain  $g^{2}=10$ obtained by turning the pumping field intensity, to realize a perfect correlation  the theoretical requirement $N_{\alpha}N_{\beta}\gg(1/g\kappa)^2=10^{19}$
and $N_{\alpha}\gg5.5N_{\beta}$ should be satisfied. Since given $N_{\beta}=10^{8}$, we choose $N_{\alpha}=10^{12}$. In the calculations, the photon numbers are much smaller than the number of atoms in the interaction region.  Thus, the density of the atomic sample is $10^{13}$-$10^{14}$/$cm^{3}$ by controlling the cell temperature. 
The signal light is turned on
after the first linear Raman splitting process and turned off right before the
second linear Raman mixing process. The pulse width of signal light should be short, and in this way, it will not affect the LRP for splitting and mixing the atomic and optical waves. Here, we set the pulse width of signal field $\sim$100 ns. To realize $N_{\beta}=10^{8}$, the power of signal light is $\sim$0.25 mW for wavelength 0.795 $\mu m$ with coherent time 100 $ns$. The mean photon number $N_{\alpha}=10^{12}$ of $\hat{a}_W^{(0)}$ field is given with  $\sim$2.5 mW of coherent time 100 $\mu s$ for wavelength 0.795 $\mu m$. The requirement for the number of photons $N_{\alpha}$ and $N_{\beta}$ can be met and be feasible in the experiment.

In conclusion, we have proposed an SU(1,1)-SU(2)-concatenated atom-light hybrid
interferometer and used it for QND measurement of photon number via the AC-Stark effect. In the scheme, the atomic spin wave of the SU(2) interferometer is prepared via a NRP and the output is detected with the method of active correlation output readout via another
NRP. Benefiting from that, the SNR in the balanced case is improved by a factor of $g$ compared to the traditional SU(2) interferometer. The condition for a perfect correlation is given. The measurement quality is reduced in the presence of losses. We
can adjust the gain parameter of the NRP in readout stage to reduce
the impact of losses. Moreover, the scheme is a multi-arm interferometer and it provides an option for the simultaneous estimation of more than two parameters 
with a wide range of applications,  such as phase imaging \cite{37}, quantum sensing networks \cite{38}, the detection of vector fields \cite{39} and so on.

\section*{Funding}

This work is supported by Development Program of China Grant No.
2016YFA0302001; National Natural Science Foundation of China Grants No.
11974111, No. 11874152, No. 91536114, No. 11574086, No. 11974116, No.
11654005; Shanghai Rising-Star Program Grant No. 16QA1401600; Innovation Program of Shanghai Municipal Education Commission No. 202101070008E00099; the Shanghai talent program, the Chinese National Youth Talent Support Program, and the Fundamental Research Funds for the Central
Universities.

\section*{Disclosures}

The authors declare no conflicts of interest.

\end{document}